\documentclass[copyright,creativecommons]{eptcs}
\usepackage{color}

\usepackage{xspace}

\usepackage{fancyvrb}
\definecolor{dkgreen}{rgb}{0,0.6,0}
\definecolor{grey}{rgb}{0.5,0.5,0.5}
\usepackage{listings}
\lstdefinelanguage{acl2}{
morekeywords={defun,defcong,defthm,t,nil,declare,defspec,local,defstructure},
keywordsprefix=:,
otherkeywords={defun-sk,rule-classes,in-theory,begin-book,include-book,in-package,mutual-recursion,instance-of-defspec},
sensitive=false,
alsodigit=-,
showstringspaces=false,
breaklines=true,
breakatwhitespace=true,
morecomment=[l];,
morecomment=[s]{\#|}{|\#},
morestring=[b]",
stringstyle=\color{dkgreen}}

\newcommand\genoc[0]{{\sc GeNoC}\xspace}
\newcommand\acl[0]{{\sc ACL2}\xspace}

\providecommand{\event}{ACL2 workshop 2013} 
\usepackage{breakurl}             
\newtheorem{definition}{Definition}

\title{A Macro for Reusing Abstract Functions and Theorems}

\author{Sebastiaan J.C. Joosten \qquad Bernard van Gastel \qquad Julien Schmaltz
\institute{Open University of the Netherlands}
\email{\{bas.joosten,Bernard.vanGastel,Julien.Schmaltz\}@ou.nl}
}

\def\titlerunning{Reusing Functions and Theorems}
\def\authorrunning{S.J.C. Joosten, B.E. van Gastel, J. Schmaltz
}
\newcommand{\ACL}[1]{\acl}
\newcommand{\iod}{\texttt{instance-of-defspec}}
\begin{document}
\maketitle

\begin{abstract}
Even though the \ACL2 logic is first order, the \ACL2 system offers several mechanisms providing users with some operations akin to higher order logic ones. In this paper, we propose a macro, named \iod{}, to ease the reuse of abstract functions and facts proven about them. \verb=Defspec= is an \ACL2 book allowing users to define constrained functions and their associated properties. It contains macros facilitating the definition of such abstract specifications and instances thereof. Currently, lemmas and theorems derived from these abstract functions are not automatically instantiated. This is exactly the purpose of our new macro. \iod{} will not only instantiate functions and theorems within a specification but also many more functions and theorems built on top of the specification. As a working example, we describe various fold functions over monoids, which we gradually built from arbitrary functions. 
\end{abstract}

\section{Introduction}


The primary goal of this paper is to provide a way to reason abstractly in \ACL2, while being able to specialise later.
This kind of reasoning is necessary in the development of large modular proofs. 
Our main contribution is a macro, called \iod{}, providing a convenient way to manipulate abstract functions and their properties. 
In our approach, we build generic theorems, and use our macro to apply them on more specific instances, that is: reuse proofs.
For this to work, we reuse functions as well, by instantiating generic functions with more specific ones.
While both function- and proof reuse are possible to some extent in \ACL2, we provide a single macro to do both in a way that is more convenient than the existing solutions.
Aside from the theoretical example presented in our paper, we briefly discuss the use of our solution
in the development of a generic theory of communication networks, called \genoc{}~\cite{Verbeek:2012cv}.
This effort is a large and modular proof development of about fifty thousands lines of \ACL2 code.
Our macro already provides a thousand lines reduction of our code base.

This paper uses monoid-operations and fold operations as leading examples.
A monoid is simply a closed and associative operation with an identity element, for which we will write $\circ$ and $0$ respectively.
A fold operation is an operation that changes a list $(a_0 \ldots a_k)$ into a value $(a_0 \circ \cdots \circ a_k)$.
The leading example is not of particular importance on its own, but used to illustrate the macro \iod{}.

Using the \verb=encapsulate= environment, \ACL2 provides a way to hide function definitions, while preserving certain function properties.
We can then prove a theorem about such functions, using only the function properties in the proof.
If we would then write a new function that satisfies all the properties used in the proof, we know that the function will satisfy the theorem as well.
However, even though we know that the function satisfies the theorem, the only way to actually use this knowledge in further proofs is by telling it to \ACL2 in the form of a new theorem.
Using an uninterpreted function in the definition of a concrete function raises similar issues.
We provide a macro called \iod{}, that will allow \ACL2 users to instantiate abstract functions while being able to use higher order theorems and functions.

The name \iod{} derives from the word \verb=defspec=, which already is a part of \ACL2~\cite{defspec}.
In essence, \verb=defspec= is a labeled \verb=encapsulate= environment, such that the functions which are kept local to the \verb=encapsulate= can be referred to as a single package.
The advantage of \verb=defspec= is that other functions can be declared to be an instance of this \verb=defspec=.
The current way of doing this in \ACL2 is by using \texttt{DefInstance}, which is a part of \verb=defspec=.
This enables us to prove the `higher order' statement that addition is a monoid.
One could then use the \texttt{:functional-instance} hint, which was already implemented in Nqthm\cite{boyer1989functional}, to prove a specific theorem if it was already proven in a more general context.
For example, we can use it to prove a theorem for addition which was already proven for a monoid efficiently.

Our approach combines these previous solutions.
At a call of \iod{}, the logical world is searched for functions and theorems that depend on the abstract \verb=defspec=.
These are copied for the concrete instance.
Using \texttt{DefInstance} and the \texttt{:functional-instance} hint, the proof of every theorem is repeated for the instance without the computational burden.

Since \iod{} will copy functions, our approach offers an alternative to \texttt{defattach}.
Kaufmann and Moore \cite{kaufmann11} explain how this can be used to add an executable function, to for instance the abstract monoid, such that we can execute it.
Unfortunately, \texttt{defattach} only allows one such attachment per function, and it does not influence the logical world.
Our macro can be seen as an extension to \texttt{DefInstance} which combines all of these solutions, making a copy of functions instead of using \texttt{defattach} thus actually applying the change in the logical world.

As a leading example for reasoning abstractly, we develop a small theory about monoids.
We prove that $(x_0 \circ (x_1 \circ (\ldots \circ (x_{n} \circ 0)))) = (((x_0 \circ x_1) \circ \ldots)\circ x_n)$ for $n\geq 0$, using any associative operation for $\circ$ with identity element $0$.
The benefit of such a proof is that we can apply it to arithmetic addition with zero, to multiplication with one, or to appending lists with the empty list.

In this paper, we present this example explaining how \iod{} can be used in Section~\ref{sec:iodUsage}.
In Section~\ref{sec:innerWorkings} we look at the inner workings of the macro, explaining limitations and opportunities in our approach.
We also describe how \iod{} can be used to add extra arguments to a function.
Similar approaches are compared to \iod{} in Section~\ref{sec:relatedwork}

\section{Monoids as an example usage}
\label{sec:iodUsage}

\begin{definition}[Monoid]
A \emph{monoid} $(0,\circ)$ on the domain $D$ is an operator $\circ$ and a constant $0$ such that:
\begin{itemize}
\item $\circ$ is closed: for all $a,b\in D$ we have $(a\circ b)\in D$.
\item $\circ$ is associative: $(a\circ b)\circ c = a\circ (b\circ c)$ for $a,b,c\in D$.
\item $0\in D$ is an identity element: $0\circ a=a=a\circ 0$.
\end{itemize}
\end{definition}
In this section, we build a monoid by adding the three constraints from the definition one at a time.
On the unrestricted operator $\circ$, we define different implementations of a fold operation.
We will show that they are equivalent for monoids.

This section is self-containing, that is: the \ACL2 interpreter should accept the inputs given in this section as-is.
For additional theorems and examples, or for less typing, the reader may use the file \verb=closedMonoid.lisp=.

First, we include our instance-of-defspec book.
\begin{lstlisting}[language=ACL2]
(include-book "instance-of-defspec")
\end{lstlisting}
This book includes an entirely abstract function called \verb=binary-function=, which was encapsulated in a \texttt{defspec} called \verb=binary=.
We repeat the definition of \verb=binary= here.
Note that this event cannot be entered in \ACL2 here, since \verb=binary= is already present in our instance-of-defspec book.
\begin{verbatim}
(defspec binary ((binary-function (x y) t))
  (local (defun binary-function (x y) (cons x y))))
\end{verbatim}
Here we use \verb=defspec=, together with \verb=local=.
The actual implementation of a \verb=local= event is unknown to \ACL2 outside the \verb=defspec=.
The only thing \ACL2 knows about \verb=binary-function= is that it takes two arguments.
Hence we can regard \verb=binary-function= as an arbitrary binary function, despite the chosen implementation of \verb=cons= here.

\subsection{Reusing functions}
The intuition of the fold operation is that it transforms a list $(x_1\ldots x_n)$ into a value $x_1 \circ \cdots \circ x_n$ for any binary operator $\circ$.
For transforming the empty list, we need some sort of identity element to build upon.
For this reason, we provide a first element $x_0$.
We define \verb=foldl= and \verb=foldr=, which differ in the placement of the brackets.
In the case of \verb=foldl=, the brackets are written as $((((x_0\circ x_1)\circ x_2)\circ x_3)\circ \cdots)$.
In the case of \verb=foldr= and \verb=foldr1=, brackets are written as: $x_1 \circ  (x_2 \circ (x_3 \circ (\cdots \circ x_0)))$.
For \verb=foldr1= we require that the list has at least one element.
For \verb=foldr= (\verb=foldl=) we supply the last (first) element.

Note that, in the end, we will prove that these fold operations all are equivalent, even when providing an identity element as the first element.
For this proof, we need associativity and an identity element, which is precisely the requirement of a monoid.

\begin{lstlisting}[language=ACL2,firstnumber=last]
(defun foldr (x xs)
  (if (atom xs) x (binary-function (car xs) (foldr x (cdr xs)))))

; Alternatively, we can 'omit' the first element:
(defun foldr1 (xs)
  (if (atom (cdr xs)) (car xs)
    (binary-function (car xs) (foldr1 (cdr xs)))))

(defun foldl (x xs)
  (if (atom xs) x
    (foldl (binary-function x (car xs)) (cdr xs))))
\end{lstlisting}

Since the focus of this article is the use of \iod{}, we do not proceed by proving properties of the fold functions, but show how to actually use these fold functions.
A trivial example of a binary function is \verb=cons=.
We can instantiate \verb=binary-function= (provided by the \texttt{defspec} encapsulate \verb=binary=) with \verb=cons= under a list of substitutions as follows:
\begin{lstlisting}[language=ACL2,firstnumber=last]
(instance-of-defspec binary cons '((binary-function cons) (foldr  cons-foldr)
                                   (foldr1 cons-foldr1)   (foldl  cons-foldl)))
\end{lstlisting}
This instantiates our fold functions as executable functions:
\begin{verbatim}
ACL2 !> (cons-foldr 'a '(b c))
(B C . A)
ACL2 !> (cons-foldr1 '(a b c))
(A B . C)
ACL2 !> (cons-foldl 'a '(b c))
((A . B) . C)
\end{verbatim}

Now lets add the assumption that our binary function is closed.
That is: there is some domain, and if both arguments to the binary function belong to it, so does its result.

We encapsulate it using a \verb=defspec=, and instantiate it as a binary operation.
We will explain the use of \verb=defspec= in Section~\ref{sec:defspec}.
You may think of it as an \verb=encapsulate= environment that hides the local definitions and provides the notion of \verb=closed-binop= in terms of \verb=c-domainp=, \verb=c-binary-function= and the theorem (to be seen as a property) \verb=closed-binop-closed=.
\begin{lstlisting}[language=ACL2,firstnumber=last]
(defspec closed-binop ((c-domainp (x) t)
                       (c-binary-function (x y) t))
  (local (defun c-domainp (x) (integerp x)))
  (local (defun c-binary-function (x y) (+ x y)))
  (defthm closed-binop-closed
    (implies (and (c-domainp x) (c-domainp y))
             (c-domainp (c-binary-function x y)))))
(instance-of-defspec binary c) ; choose c (for closed) as the prefix symbol here
\end{lstlisting} \label{sec:domainp}
In the last statement, we instantiate \verb=closed-binop= as an arbitrary binary operator: we use abstract functions as the instantiation of other abstract functions.
Even though a closed binary function is an abstraction itself, it is an instantiation of the - more general - binary function.

Note that in this case, we did not specify the list of replacements.
We do not have to: the second argument \verb=c= is used as the default prefix.
Hence, \verb=c-foldr= is the instantiation of \verb=foldr= with \verb=c-binary-function= as \verb=binary-function=:
\begin{verbatim}
ACL2 !> :pf c-foldr
(EQUAL (C-FOLDR X XS)
       (IF (CONSP XS)
           (C-BINARY-FUNCTION (CAR XS)
                              (C-FOLDR X (CDR XS)))
           X))
\end{verbatim}
In fact, we could have done the same with \verb=cons=.
The following would have been a shorter notation for the same instruction we gave earlier:
\begin{lstlisting}[language=ACL2,firstnumber=last]
(instance-of-defspec binary cons '((binary-function cons)))
\end{lstlisting}

\subsection{Reusing theorems}
We have seen how to reuse functions using \verb=defspec=.
We can do the same trick for theorems.
In the context of a closed operation, we show that the repetitive application of this operation again yields an element in its domain.
For brevity, we only show this for the \verb=foldr1= operation:
\begin{lstlisting}[language=ACL2,firstnumber=last]
(defun list-domainp (xs)
  (if (endp xs) t
    (and (c-domainp (car xs)) (list-domainp (cdr xs)))))

(defthm foldr1-closed
  (implies (and (list-domainp xs) (consp xs))
           (c-domainp (c-foldr1 xs))))
\end{lstlisting} \label{sec:foldr1-closed}

We reuse this theorem for a semigroup.
A closed associative operator $\circ$ is called a semigroup.
Note that a semigroup is like a monoid, but without the identity element $0$.
More formally: a monoid $(0,\circ)$ is a semigroup $\circ$ with identity element $0$.
So once again, we write a \texttt{defspec}, and specify that it is an instance of a closed binary operator.
\begin{lstlisting}[language=ACL2,firstnumber=last]
(defspec semigroup ((sg-c-domainp (x) t)
                    (sg-c-binary-function (x y) t))
  (local (defun sg-c-domainp (x) (integerp x)))
  (local (defun sg-c-binary-function (x y) (+ x y)))
  (is-a closed-binop sg semigroup-is-a-closed-binop)
  (defthm semigroup-assoc
    (implies (and (sg-c-domainp x)
                  (sg-c-domainp y)
                  (sg-c-domainp z))
             (equal (sg-c-binary-function x (sg-c-binary-function y z))
                    (sg-c-binary-function (sg-c-binary-function x y) z)))))
(instance-of-defspec closed-binop sg) ; reuse the fold operators (again)
\end{lstlisting}

Note that we used the macro \verb=is-a= inside our semigroup.
This copies the theorems from the closed binary operator into the current \texttt{defspec}.
By doing so, we ensure that the previously defined specification of \verb=closed-binop= will be copied to \verb=semigroup=, since this is required to prove that a \verb=semigroup= is an instance of \verb=closed-binop=.
The \verb=is-a= macro is auxiliary to \iod{}. Its implementation will be discussed in Section~\ref{sec:is-a}.
In this case, the following theorem is generated in place of the \verb=is-a= macro:
\begin{verbatim}
ACL2 !> (OLDSPEC 'CLOSED-BINOP 'SEMIGROUP-IS-A-CLOSED-BINOP 'SG () (W STATE))
((DEFTHM SEMIGROUP-IS-A-CLOSED-BINOP-0
         (IMPLIES (IF (SG-C-DOMAINP X)
                      (SG-C-DOMAINP Y)
                      'NIL)
                  (SG-C-DOMAINP (SG-C-BINARY-FUNCTION X Y)))))
\end{verbatim}

Now we try to prove that \verb=foldr1= is \verb=foldl=.
Note that this was not the case for \verb=cons-foldr1= and \verb=cons-foldl=.
We need to prove it for the semigroup folds: \verb=sg-c-foldr1= and \verb=sg-c-foldl=.
\begin{lstlisting}[language=ACL2,firstnumber=last]
(defthm foldr1-is-foldl
  (implies (and (sg-c-domainp x) (sg-c-domainp y)
                (sg-list-domainp xs))
           (equal (sg-c-foldr1 (cons x xs))
                  (sg-c-foldl x xs))))
\end{lstlisting}

If we look at the proof output, we will find that the proof has used a theorem we did not define ourselves,
but which was automatically copied based on the theorem we added to the closed binary operator:
\begin{verbatim}
ACL2 !> :pf (:REWRITE SG-FOLDR1-CLOSED)
(IMPLIES (AND (SG-LIST-DOMAINP XS) (CONSP XS))
         (SG-C-DOMAINP (SG-C-FOLDR1 XS)))
\end{verbatim}
Without the theorem \verb=foldr1-closed=, and thus the automatically derived theorem \verb=sg-foldr1-closed=, the proof attempt of \verb=foldr1-is-foldl= would have failed.

\subsection{On monoids}
As promised, we end with a theory about monoids.
A monoid is a semigroup with an identity element.
To take care of the names, we use a renaming constant.
Note that we can use renaming in the macro \verb=is-a=, exactly like in \iod{}.
\begin{lstlisting}[language=ACL2,firstnumber=last]
(defconst *monoid-renaming*
  '((sg-c-domainp         mon-domainp)      (sg-c-foldr  mon-foldr)
    (sg-c-binary-function mon-binop)        (sg-c-foldr1 mon-foldr1) 
    (sg-list-domainp      mon-list-domainp) (sg-c-foldl  mon-foldl) ))

(defspec monoid ((mon-domainp (x) t) (mon-binop (x y) t)
                 (mon-id () t))
  (local (defun mon-domainp (x) (integerp x)))
  (local (defun mon-binop (x y) (+ x y)))
  (local (defun mon-id () 0))
  (defthm id-in-domain (mon-domainp (mon-id)))
  (is-a semigroup mon monoid-is-a-semigroup *monoid-renaming*)
  (defthm monoid-id-left
    (implies (and (mon-domainp x))
             (equal (mon-binop x (mon-id))
                    x)))
  (defthm monoid-id-right
    (implies (and (mon-domainp x))
             (equal (mon-binop (mon-id) x)
                    x))))
(instance-of-defspec semigroup mon *monoid-renaming*)
\end{lstlisting}

We introduce function \verb=fold= which acts like \verb=foldr1=, but without the requirement that its argument should be a \verb=consp=.
\begin{lstlisting}[language=ACL2,firstnumber=last]
(defun fold (xs) (if (atom xs) (mon-id) (mon-foldr1 xs)))
\end{lstlisting}
We end the book about monoids, the file \verb=closedMonoid.lisp=, by proving equality between the different versions of \verb=fold=, and giving another instantiation.
These proofs are omitted here, but we encourage the reader to take a look.

\section{Inner workings}
\label{sec:innerWorkings}
As the sources will be made available with this publication, we do not reproduce them here.
Instead, we highlight the main parts to give the reader a rough understanding of the code, and highlight the `ugly bits' to illustrate the difficulties and limitations of our approach.

The macro \iod{} effectively does three things:
\begin{enumerate}
\item Look up the \texttt{defspec} and prove that the provided instance is an instance of this \texttt{defspec} using \texttt{DefInstance}.
\item Look up every function that uses one of the \texttt{defspec} functions, and copy it as an instantiated function.\label{step:functions}
\item Look up every theorem that use one of the functions from the previous step, and copy it as an instantiation.
\end{enumerate}

In order to find the functions and theorems, we look at the \ACL2 \texttt{world}.
For this reason, we use make-event.
In particular, \iod{} is a macro that expands to:
\begin{verbatim}
`(make-event (instanceOf-defspec-fn ',spec ',prefix ,rename state)
  :check-expansion nil)
\end{verbatim}

\subsection{Obtaining the \texttt{defspec}}\label{sec:defspec}
To reason abstractly, \ACL2 provides the \verb=encapsulate= environment.
This environment allows the user to hide events, such that a particular theorem can be stated about a function, without allowing the definition of this function to enter the logical world.
When proving further properties outside the encapsulate, the function is seen as an abstract function, since the function is unknown to the logical world.

To prove that a certain concrete function is an instance of this environment, \verb=defspec= was developed by Sandip Ray and Matt Kaufmann.
The system book \verb=make-event/defspec.lisp= provides two parts:
the first is a macro called \verb=defspec=, which does exactly the same as an \verb=encapsulate= environment, but then also provides a name for it.
The second is a macro called \verb=definstance=, which generates a theorem equivalent to stating that some implementation is an actual instantiation of the \verb=defspec= named.
We chose to build on this approach for the reason why Ray and Kaufmann developed it.
In the comments of their code, they write~\cite{defspec}:
\begin{quote}
The real problem is that ACL2 is a theorem prover for first order logic, not
higher-order logic, while the statement we want to make is inherently a
higher-order statement. (...)

But having one macro that generates the instances will give the evaluators the
ability to trust it, rather than hand-coded proofs that all the ``corresponding"
constraints are satisfied.  The hope is that with a lot of use the macros here
will be conventionally thought of as higher-order representations.
\end{quote}

To obtain the \verb=defspec=, we use \verb=acl2::decode-logical-name= to skip to the place where the defspec was declared, and lookup the corresponding encapsulate event.
In the case of \verb=closed-binop= as written in the previous section, it might look like this:
\begin{verbatim}
 (EVENT-LANDMARK GLOBAL-VALUE 8844
                 (ENCAPSULATE (C-DOMAINP C-BINARY-FUNCTION)
                              . :COMMON-LISP-COMPLIANT)
                 ENCAPSULATE
                 ((C-DOMAINP (X) T)
                  (C-BINARY-FUNCTION (X Y) T))
                 (LOCAL (DEFUN C-DOMAINP (X) (INTEGERP X)))
                 (LOCAL (DEFUN C-BINARY-FUNCTION (X Y) (+ X Y)))
                 (DEFTHM CLOSED-BINOP-CLOSED
                         (IMPLIES (AND (C-DOMAINP X) (C-DOMAINP Y))
                                  (C-DOMAINP (C-BINARY-FUNCTION X Y)))))
\end{verbatim}
From this, we obtain the corresponding functions by taking the \verb=cadar= of the third element.
In this case, it returns \verb=(C-DOMAINP C-BINARY-FUNCTION)=.
At this point, we would like to note that we do not know whether this is an appropriate way to find the encapsulated functions.
We just chose this to identify the \verb=defspec= because it consistently returned these functions in multiple tests.

Instantiating the \verb=defspec= is done with a \verb=definstance=.
At the instantiation of \verb=closed-binop= as a semigroup, we generated the following statement:
\begin{verbatim}
(DEFINSTANCE CLOSED-BINOP SG-CLOSED-BINOP
       :FUNCTIONAL-SUBSTITUTION ((C-DOMAINP SG-C-DOMAINP)
                                 (C-BINARY-FUNCTION SG-C-BINARY-FUNCTION)))
\end{verbatim}
This is the first statement generated.
It is also the main proof obligation: it will fail when \ACL2 cannot prove that the instance presented is an implementation of the \verb=defspec=.

\subsection{Obtaining functions}
Once we have found the functions defined in the \verb=defspec=, we look up all functions that depend on these.
Note that we apply this transitively.
That is: if we use the abstract function $f$ inside $g$, and $g$ is used inside $h$, then $h$ has to be instantiated as well.
To do so, we wrote function \verb=get-derived-funs=.

Function \verb=get-derived-funs= goes through the \texttt{world} multiple times, keeping track of a list of `discovered' functions while looking for functions that use one of these 
and adding them to the list.
The search for functions is done by looking for \verb=DEF-BODIES= in the \texttt{world}, which ensures that macros have been eliminated from the definition.
We terminate this search once the list does not grow any further.
For each separate function, we stop searching once we hit the definition of that function.
The rationale behind this is that a function cannot be used before it was declared.
Since the procedure will just be used in a \verb=make-event=, we can leave it in \verb=:program= mode, saving us the trouble to prove termination.
We take care not to add functions twice, which does provide termination: there are a finite number of functions in the current logical world.

When copying the function, we wish to be as similar to the original function as possible.
However, we have to replace the abstract functions for their instantiations.
To do so, we expand all macros in the function body using the \verb=:trans= macro.
After this, we replace the functions with their definitions using a function we called \verb=replacefns=.
This function takes a list of desired substitutions, and a list of terms in which this replacement should take place.
\begin{verbatim}
ACL2 !> :replacefns ((foo bar) (bar foo))
               ((+ ((lambda (foo j) (foo foo j)) x y) (bar x y)))
((+ ((LAMBDA (FOO J) (BAR FOO J)) X Y) (FOO X Y)))
\end{verbatim}

Copying our function as identically as possible has the advantage of copying documentation and other parameters as well.
There are cases where the generated \verb=defun= fails, which occur when \ACL2 fails to prove guards or termination of the generated function.

\subsection{Obtaining theorems}
Obtaining the theorems happens in a similar way.
The main difference is that the list of functions does not grow while searching for theorems that use these functions.
We can therefore find all theorems in one pass through the \texttt{world}.

When generating the theorems, we do not try to let the copied theorem mimic the original, as we did with functions.
In the case of functions, it was possible that \ACL2 tried to prove something (like guards or termination), but failed.
In the case of theorems, we want to prevent this from happening, by making sure that \ACL2 does not redo an entire proof.
To achieve this, we use the \verb=:functional-instance= hint.
Also, we do not try to copy the original event, but look for its effect on the logical world and copy the effect.

The theorem \verb=foldr1-closed= from Section~\ref{sec:foldr1-closed} is stored in the \texttt{world} as:
\begin{verbatim}
(...
 (FOLDR1-CLOSED THEOREM IMPLIES
                (IF (LIST-DOMAINP XS) (CONSP XS) 'NIL)
                (C-DOMAINP (C-FOLDR1 XS))) ...
 (FOLDR1-CLOSED CLASSES (:REWRITE))
 ...)
\end{verbatim}
Note that the theorem and the classes are stored in different \texttt{world} items.
We obtain the theorems by combining these two parts in the \texttt{world}.
The first holds the (translated, macro-free) theorem, and the second holds the rule-class(es).
Some rule-classes are stored with extra attributes.
For example, a typing rule will define a typed term, and a forward-chaining rule has a trigger-term.

The \verb=:functional-instance= hint will generate various subgoals, which can be proven using the proof generated by a previously proven \verb=definstance= theorem (which is the first theorem we generate, even before copying the functions).
We also add the \verb=:in-theory= hint, giving a theory that contains only the newly generated functions.
Although we do not know how to prove that this is a sound way to copy theorems, we can at least give the anecdotical evidence that we have not found an example where the proof attempt for a generated theorem failed.

While developing \iod{}, we found it useful to be able to keep track of the theorems defined.
With relatively little effort, we have written the macro \verb=symbol-lemmas= which, like \verb=:pl=, shows theorems containing its argument.
The main difference being that the former shows all theorems, while the latter only shows those in which its argument is a trigger.

\subsection{Adding arguments}\label{sec:addingarguments}
The main use of \iod{} is to provide a single method to create functions like \verb=map=, and corresponding theorems.
An issue we did not foresee was that some of the functions we would like to map, though effectively unary, are actually binary.

We have solved this by allowing for lambda functions in the function substitution.
As an example, we investigate a generic function that checks whether all elements in some list satisfy some predicate:
\begin{lstlisting}[language=ACL2,numbers=none]
(defspec list-predicate ((predicate (x) t))
  (local (defun predicate (x) x)))
(defun predicate-listp (lst)
  (if (atom lst) (null lst) ; require true-lists
    (and (predicate (car lst)) (predicate-listp (cdr lst)))))
\end{lstlisting}
Now suppose we want our predicate to be \verb=member-equal=.
This would enable us to test whether all elements in some list occur in some other list, which is exactly the condition for subsets.
We can create a new function \verb=subset-equal= (written without a \verb=p= to prevent collision with the builtin \verb=subsetp-equal=):
\begin{lstlisting}[language=ACL2,numbers=none]
(instance-of-defspec list-predicate members
  '((predicate (lambda (x) (member-equal x y)))
    (predicate-listp (lambda (lst) (subset-equal lst y)))))
\end{lstlisting}
At this point, we require the lambda expression to have exactly the same argument as the arguments of the function it replaces:
\begin{verbatim}
ACL2 !> (instance-of-defspec list-predicate members
          '((predicate (lambda (x) (member-equal x y)))
            (predicate-listp (lambda (xs) (subset-equal xs y)))))

ACL2 Error in COPYFUN:  The lambda construct 
(LAMBDA (XS) (SUBSET-EQUAL XS Y)) takes as input (XS), which should
be an exact match of the original arguments of the original function:
(LST)
\end{verbatim}
This syntactic limitation will hopefully prevent users unintentionally swapping arguments, and helps identify which variables are substituted.

\subsection{The \texttt{is-a} macro}\label{sec:is-a}
To prove something is an instance of some \verb=defspec=, we need to prove all theorems included in that \verb=defspec=.
If we want to add a property to a previous \verb=defspec=, we proceed by creating a new \verb=defspec=, and indicating that the new is an instance of the former.
One way to do this, is by using \verb=definstance=, and another is by using \verb=is-a=.
The main advantage of the latter is that theorems are be copied individually, and can remain enabled.
This results in a different behaviour of subsequent proof attempts.
Another advantage is that the notation of \verb=is-a= is very similar to that of \iod{}.

The implementation of \verb=is-a= is a lot like \verb=definstance=, in the sense that it uses the function \verb=constraint= defined in the \verb=defspec= book to get all theorems that have to hold.
For every function it finds in the \verb=defspec=, the function \verb=constraint= is called which returns a theorem that must hold.
While it would be possible to use \verb=is-a= outside of a \verb=defspec=, it would make more sense to use \iod{} there.

\section{Discussion}
\paragraph{Related work}\label{sec:relatedwork}
Before we started working with \iod{}, we used \ACL2's macro system to avoid code duplication.
This approach does not involve inspecting the world, or \verb=make-event=.
We just used plain macros that write out `instances'.
Apart from functions, we used these macro's to generate theorems as well.
In this approach we lose the proof obligation that the proposed instance is actually an instance, or even having to create an encapsulated environment for the abstract structure.
The price we pay, however, is having to do full proofs for all instantiated theorems, which can be a heavy computational burden.
Also, we found the current approach of writing theorems about abstract instances directly in the logical world to be more convenient than writing theorems in a macro.
Carl Eastlund and Matthias Felleisen make a case against using \ACL2's macros, proposing hygienic macros~\cite{Eastlund11}.
If we were to use hygienic macros for this purpose, writing out functions and theorems using a macro could be more convenient.
We would still need to redo theorems.

A very promising alternative to our approach was presented by Gamboa and Patterson~\cite{gamboa03}.
The authors introduce a way to use polymorphism in \ACL2.
One of the downsides is the use of a \verb=stobj= called \verb=memory=.
In addition, it is presented as an alternative to the current \verb=encapsulate= environments, instead of building on what is already present in \ACL2.

Although we could not find a paper on it, there is a macro called \verb=def-functional-instance= inside the \ACL2 tools directory which provides an easier way to instantiate previous theorems using the \verb=:functional-instance= hint.
The problem is solved more rigorously by Moore~\cite{moore09}, by providing an alternative to the \verb=:functional-instance= hint which computes its substitutions automatically.
Our \iod{} macro does all this, and these two previous solutions stress the importance of doing so.

For conveniently instantiating functions, the macro by Mart\'in-Mateos et al.~\cite{alonso02} is most similar to ours.
The same instantiations are performed: both functions and theories can be reused.
The authors did not use the \verb=defspec= environment, nor do their macros rewrite proof hints to aid the instantiated proofs.
Also, at some point in the process the user is required to specify exactly which events to copy (these must be \verb=defun= and \verb=defthm= events).
A macro call must be placed around these events, and the output of this macro should be copied manually.
The user then obtains a program which will define a constant that can be used for reuse.
One of the reasons that their approach was less convenient than ours, is that \ACL2 did not have some of its current features, especially \verb=make-event=, at the time.
Indeed, such improvements are present in later work.
In the code accompanying the more recent paper from this group~\cite{Lamban12}, many of the improvements suggested in the earlier paper~\cite{alonso02} have been made.
Fundamental differences still are:
\begin{itemize}
\item The macro by Mart\'in-Mateos only copies a handful of events, where our macro will take any event that rewrites itself to a \verb=defun= or a \verb=defthm=, which includes those handled by Mart\'in-Mateos.
\item The macro by Mart\'in-Mateos requires the general theory to be wrapped inside a macro entirely. Our approach only requires this for the local \verb=defspec=, allowing the general theory to be distributed over multiple books.
\item Our macro allows arguments to be added to function calls without changing the original function.
\item Leaving the definitions of the instantiation enabled might cause the macro by Mart\'in-Mateos to fail. In our approach the theorem prover is guided by automatically generated hints. Since both approaches use the \verb=:functional-instance= hint, we expect this to be something which can be resolved rather easily.
\end{itemize}

Carl Eastlund and Matthias Felleisen built a module system on top of \ACL2, using the Dracula environment~\cite{Eastlund09}.
In this system, the abstract entities are called interfaces, which have signatures and contracts.
This compares well to the \verb=defspec= which is already built in \ACL2.
A major difference is that, in \ACL2, a function has to be declared \verb=local=, and the theorems must hold for it.
In essence, the user must provide a witness for the \verb=defspec=, while the Dracula environment does not require this.
The interfaces can be instantiated as modules, or reused in modules, by using export and import commands respectively.
An advantage of Dracula Modular ACL is that modules are checked independently, which might improve the performance of \ACL2 by reducing memory requirements.
For our approach, one would have to write different files in order to get such behavior.
We see two disadvantages to the modular approach.
\begin{itemize}
\item Reasoning about interfaces directly is not possible.
A familiar proof states that there can only be one identity element for a monoid.
In the modular approach, one would have to create a module which imports the interface, and exports another interface which also has the property that there is only one identity element.
In their paper, Eastlund and Felleisen prove correctness of the executable modules.
Hence the module that describes a property about the monoid interface will not be proven to be correct until a witness for the monoid is provided.
\item The Dracula environment requires the user to trust both the \ACL2 system and the module system, whereas in our approach all proofs are performed within the \ACL2 system.
Although we believe the module system to be sound, bugs in \ACL2 up to version 2.3 show that reasoning with encapsulated properties is particularly error-prone.
The reader may, for instance, read the discussion on subversive recursions in the miscellaneous section of the \ACL2 documentation (version 3 and up).
\end{itemize}

\paragraph{Use in practice}
Our group is involved in a large scale generic proof about Network-On-Chips, called \genoc{}~\cite{Verbeek:2012cv}.
The proofs consists of almost fifty thousand lines of \acl code.
In \genoc{} all kinds of properties about classes of networks are proven.
To maximise proof reusability, an important aspect is partially instantiating these generic proofs for classes of networks, and later fully instantiating the proofs for concrete instantiations of networks.
Currently we use many hand coded functional substitution rules in the \verb=definstance=.
These are hard and error-prone to maintain.
By using the \iod{} construct we already reduce our effort to maintain our codebase.
Also a lot of instantiations of generic theorems can be removed, because they are automatically generated.
Both these changes reduced our code base by one thousand lines.

As \iod{} copies every function and theorem based on a \verb=defspec=.
Because of this, instantiating a defspec early in a proof session and instantiating it late in the session could have different meanings.
For this reason, we ensured that it would be possible to instantiate the same set of functions as a defspec twice.
One might worry that all extra instantiated definitions can clutter up the logical world and namespace.
In practice, however, we only needed to instantiate each set of functions as one defspec, and did not run into problems with a cluttered up namespace.
Nevertheless we provided a feature, per request by one of the reviewers, which allows you to not copy some theorems.
To do this, add a theorem to the rename list without providing a new name for it.
For instance, if \verb=sometheorem= would normally be instantiated by a call to \iod{}, add \verb=(sometheorem newname)= to use \verb=newname= as the new name for this theorem, or \verb=(sometheorem)= to not copy \verb=sometheorem= at all.

\paragraph{Future work}
Using our macro in \emph{all} instances in GeNoC still presents a problem.
We cannot handle function definitions inside \verb=encapsulate= environments correctly.
To find functions to use as \verb=local= witness, we need to look inside the \verb=encapsulate= environment, which is forbidden.
For this reason, the \verb=defun-sk= and \verb=defevaluator= events, which rewrite to events including the \verb=encapsulate=, are not supported.

In the current example, we use \verb=is-a= to include a previous \verb=defspec= into a new one.
In practice, not all functions may be duplicated in this step.

We can already illustrate how this problem arises by a modification in the monoid example.
Suppose that in Section~\ref{sec:domainp}, instead of adding the \verb=domainp= function to \verb=closed-binop=, we would have used \verb=predicate= and the corresponding generic function \verb=predicate-listp= (possibly included from a different file).
Proving that \verb=closed-binop= is an instance of \verb=binary= will not be a problem.
Proving that the \verb=semigroup= is an instance of \verb=closed-binop=, however, requires instantiating \verb=predicate= with \verb=sg-c-domainp= \emph{while} instantiating the semigroup operator with \verb=closed-binop=.
We are in the process of finding a convenient syntax for this, and we will provide a way to do this in the near future.

\section{Conclusions}
The macro \iod{} enables us to reuse abstract functions and facts proven about them.
It automatically instantiates lemmas and theorems derived from the abstract functions.
We are convinced that together with \verb=defspec=, our macro provides a way to reason abstractly in \ACL2, while being able to specialise later.
The great variety of similar approaches shows that there is a strong need for abstract reasoning in \ACL2. This variety 
of solutions also suggests that no approach has become the standard for the community. 
We believe that our approach is more convenient than the existing solutions, and we hope it will become a standard in the \ACL2 community.

\paragraph{Acknowledgments}
We thank the reviewers for their detailed comments and apposite insights.

\bibliographystyle{eptcs}
\bibliography{generic}
\end{document}